# Calibrating photon counts from a single image

Rainer Heintzmann[1,2*], Peter K. Relich[3], Robert P.J. Nieuwenhuizen[4],
Keith A. Lidke[3] and Bernd Rieger[4*]

[1]Leibniz Institute of Photonic Technology, Jena, Germany
[2]Institute of Physical Chemistry and Abbe Center of Photonics, Friedrich Schiller University, Jena, Germany
[3]Department of Physics and Astronomy, University of New Mexico, Albuquerque, NM, USA
[4]Department of Imaging Physics, Delft University of Technology, Delft, The Netherlands
* Correspondence to: heintzmann@gmail.com or b.rieger@tudelft.nl

## Abstract

**Current methods for detector gain calibration require acquisition of tens of special calibration images. Here we propose a method that obtains the gain from the actual image for which the photon count is desired by quantifying out-of-band information. We show on simulation and experimental data that our much simpler procedure, which can be retroactively applied to any image, is comparable in precision to traditional gain calibration procedures.**

Optical recordings consist of detected photons, which typically arrive in an uncorrelated manner at the detector. Therefore the recorded intensity follows a Poisson distribution, where the variance of the photon count is equal to its mean. In many applications images must be further processed based on these statistics and it is therefore of great importance to be able to relate measured values $S$ in analogue-to-digital-units (ADU) to the detected (effective) photon numbers $N$. The relation between the measured signal $S$ in ADU and the photon count $N$ is given by the linear gain $g$ as $S = gN$. Only after conversion to photons is it possible to establish the expected precision of intensities in the image, which is essential for single particle localization, maximum-likelihood image deconvolution or denoising [Ober2004, Smith2010, Afanasyev2015, Strohl2015]. The photon count must be established via gain calibration, as most image capturing devices do not directly report the number of detected photons, but a value proportional to the photoelectron charge produced in a photomultiplier tube or collected in a camera pixel. For this calibration typically tens of calibration images are recorded and the linear relationship between mean intensity and its variance is exploited [vanVliet1998]. In current microscopy practise a detector calibration to photon counts is often not done but cannot be performed in retrospect. It thus would be extremely helpful, if that can be determined from analysing the acquisition itself – a single image. A number of algorithms have been published for Gaussian type noise [Donoho1995, Immerkaer1996] and Poissonian type noise [Foi2008, Colom2014, Azzari2014, Pyatykh2014]. However, all these routines use assumed image properties to extract the information rather than just the properties of the acquisition process as in our presented algorithm. This has major implications for their performance on microscopy images governed by photon statistics (see Supplementary Information for a comparison with implementations from Pyatykh et al. [Pyatykh2014] and Azzari et al. [Azzari2014] which performed more than an order of magnitude worse than our method).

Some devices, such as avalanche photodiodes, photomultiplier tubes (PMTs) or emCCD cameras can be operated in a single photon counting mode [Chao2013] where the gain is known to be one. In many cases, however, the gain is unknown and/or a device setting. For example, the gain of PMTs can be continuously controlled by changing the voltage between the dynodes and the gain of cameras may deviate from the value stated in the manual. To complicate matters, devices not running in photon counting mode, use an offset $O_{zero}$ to avoid negative readout values, i.e. the device will yield a non-zero mean value even if no light reaches the detector, $S = gN + O_{zero}$. This offset value $O_{zero}$ is sometimes changing over time ("offset drift"). Traditionally, a series of about 20 dark images and 20 images of a sample with smoothly changing intensity are recorded [vanVliet1998]. From these images the gain is calculated as the linear slope of the variance over these images versus the mean intensity $g = \text{var}(S)/\text{mean}(S)$ (for details see Supplementary information). In **Figure 1** we show a typical calibration curve by fitting (blue line) the experimentally obtained data (blue crosses). The obtained gain does not necessarily correspond to the real gain per detected photon, since it includes multiplicative noise sources such as multiplicative amplification noise, gain fluctuations or the excess noise of emCCDs and PMTs. In addition there is also readout noise, which includes thermal noise build-up and clock induced charge. The unknown readout noise and offset may seem at first glance disadvantageous regarding an automatic quantification. However, as shown below, these details do not matter for the purpose of predicting the correct noise from a measured signal.

Let us first assume that we know the offset $O_{zero}$ and any potential readout noise variance $V_{read}$. The region in Fourier space above the cut-off frequency of the support of the optical transfer function only contains noise in an image [Liu2017], where both Poisson and Gaussian noise are evenly distributed over all frequencies [Chanran1990, Liu2017]. By measuring the spectral power density of the noise $V_{HF}$ in this high-frequency out-of-band region and accounting for the area fraction $f$ of this region in Fourier space, we can estimate the total variance $V_{all}=V_{HF}/f$ of all detected photons. Then the gain $g$ is then obtained as

(1) $$g = \frac{V_{all}-V_{read}}{\sum(S-O_{zero})}$$

where we relate the photon-noise-only variance $V_{all}$-$V_{read}$ to the sum offset-corrected signal over all pixels in the image (see Online Methods). The device manufacturers usually provide the readout noise leaving only the offset and gain to be determined from the image itself in practise. To also estimate both, the offset together with the gain, we need more information from the linear mean-variance dependence than given by equation (1). We achieve this by tiling the input image, e.g. into 3×3 sub-images, and process each of these sub-images to generate one data point in a mean-variance plot. From these nine data points we obtain the axis offset ($O_{no-noise}$). We then perform the gain estimation (1) on the whole image after offset correction (See Online Methods and Supplementary Information). As seen from **Figure 1** the linear regression of the mean-variance curve determines the axis offset ADU value $O_{no-noise}$ at which zero noise would be expected. Yet we cannot simultaneously determine both offset $O_{zero}$ and readout noise $V_{read}$. If either of them is known a priori, the other can be calculated: $V_{read} = g(O_{zero} - O_{no-noise})$, which is, however, not needed to predict the correct noise level for each brightness level based on the automatically determined value $O_{no-noise}$.

To test the single-image gain calibration, simulated data was generated for a range of gains (0.1, 1, 10) with a constant offset (100 ADU), a range of readout noise (1, 2, 10 photon RMS) and maximum expected photon counts per pixel (10, 100, …, $10^5$). Offset and gain were both determined from band-limited single images of two different objects (resolution target and Einstein) without significant relative errors in the offset or gain (less than 2% at more than 10 expected maximum photon counts) using the proposed method (see **Supplementary Figures S1-S3**). **Figure 1** quantitatively compares the intensity dependent variance predicted by applying our method individually to many single experimental in-focus images (shaded green area) with the classical method evaluating a whole series of calibration images (blue line). Note that our single-image based noise determination does not require any prior knowledge about offset or readout noise. **Figure 2** shows a few typical example images acquired with various detectors together with the gain and offset determined from each of them and the calibration values obtained from the standard procedure for comparison. We evaluated the general applicability of our method on datasets from different detectors and modes of acquisition (CCD, emCCD, sCMOS, PMT, GAsP and Hybrid Detector). **Figure 3** quantitatively compares experimental single image calibration with classical calibration. 20 individual images were each submitted to our algorithm and the determined offset and gain was compared to the classical method. The variance of a separately acquired dark image was submitted to the algorithm as a readout noise estimate, but alternatively the readout noise specification from the handbook could be used or a measured offset at zero intensity. As seen from **Figure 3** the single-image-based gain calibration as proposed performs nearly as well as the standard gain calibration using 20 images. The relative gain error stays generally well below 10% and for cameras below 2%. The 8.5% bias for the HyD photon counting system is unusually high, and we were unable to find a clear explanation for this deviation from the classical calibration. Using only lower frequencies to estimate $V_{HF}$ ($k_t$ =0.4) resulted in a much smaller error of 2.5% in the single-photon counting case suggesting that dead-time effects of the detector might have affected the high spatial frequencies.

Simulations as well as experiments show a good agreement of the determined gain with the ground truth or gold standard calibration respectively. The bias of the gain determined by the single-image routine stayed below 4% (except for HyD). For intensity quantification any potential offset must be subtracted before conversion to photon counts. Our method estimates the photon count very precisely over a large range of parameters (relative error below 2% in simulations). Our method could be applied to many different microscopy modes (widefield transmission, fluorescence, and confocal) and detector types (CCD, emCCD, sCMOS, PMT, GAsP and HyD photon counting), because we only require the existence of an out-of-band region, which purely contains frequency independent noise. This is usually true, if the image is sampled correctly. As discussed in the Supplementary Information the cut-off limit of our algorithm can in practise be set below the transfer limit and single-image calibration can even outperform the standard calibration if molecular blinking perturbs the measurement

In summary we showed that single image calibration is a simple and versatile tool. We expect our work to lead to a better ability to quantify intensities in general. This will have significant impact on all methods that profit from knowledge about the noise distribution. Examples range from quantification of light intensities over parameter selection for patch-based denoising and deconvolution procedures to noise prediction.

## Author Contribution

RH conceived the initial idea. BR and RH extended it and wrote the initial software. PR performed experiments. BR and RH performed simulations and wrote the paper. PR and RPJN contributed to solve the offset estimation. KAL gave research advice. All authors commented on the paper.

## Competing financial interests

The authors declare to have no competing interest.

## Acknowledgements

We thank Steffen Dietzel, Core Facility Bioimaging LMU München, for providing calibration datasets. Colin Park and Simon Denham, Carl Zeiss U.K., are thanked for their help in measuring data on the Zeiss LSM 800. Ute Neugebauer and Christina Große are thanked for help in taking data on a Zeiss 800 system using ordinary PMTs. Many thanks to Rüdiger Bader (Hamamatsu) for his support in the sCMOS measurements. Gabriela Ficz is thanked for discussion in initial stages of the manuscript. Diederik Feilzer is thanked for helping coding the ImageJ plugin. P.K.R. and K.A.L. were supported by National Science Foundation grant no. 0954836. R.H. acknowledges Deutsche Forschungsgemeinschaft grants TRR 166 TP A4, B5 and HE3492/7-1. B.R. acknowledges European Research Council grant no. 648580 and National Institute of Health grant no. 1U01EB021238-01.

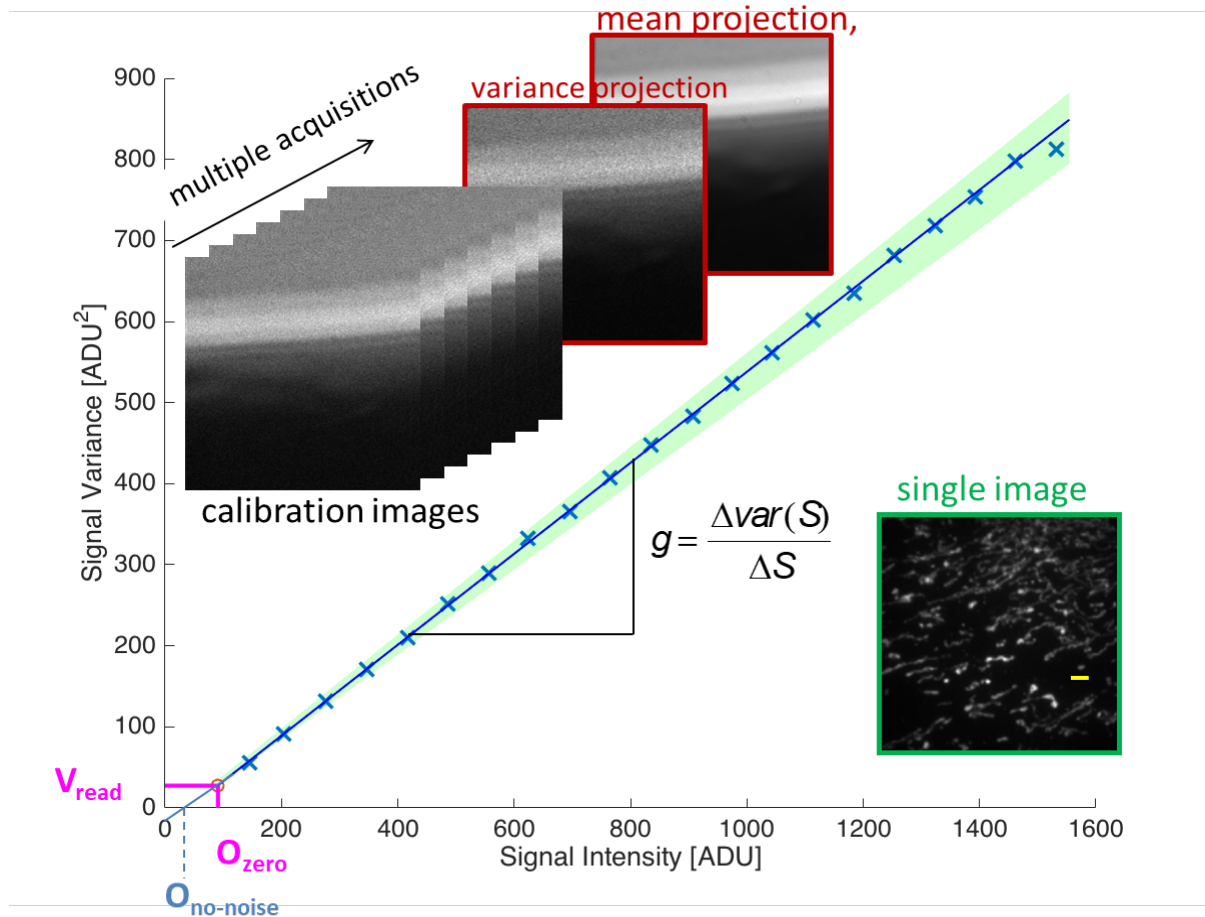

**Figure 1. Gain estimation by plotting variance over mean.** Typical detector calibration curve obtained from a series of static images. A series of 20 images of a defocussed edge (left inset) and a background series of 20 images are processed into a variance and mean projection image. From these a mean-variance intensity plot is obtained (only 21 data points are displayed as blue crosses for better visualization). The experimentally found mean background ($O_{zero}$) and the readout noise $V_{read}$, the variance at zero intensity, are indicated. The gain $g$ is obtained as the slope of the linear regression (blue line). The shaded green area corresponds to the minimum and maximum values for single image gain and offset results obtained from 20 single in-focus images with our method (bottom right inset, scale bar 2 µm).

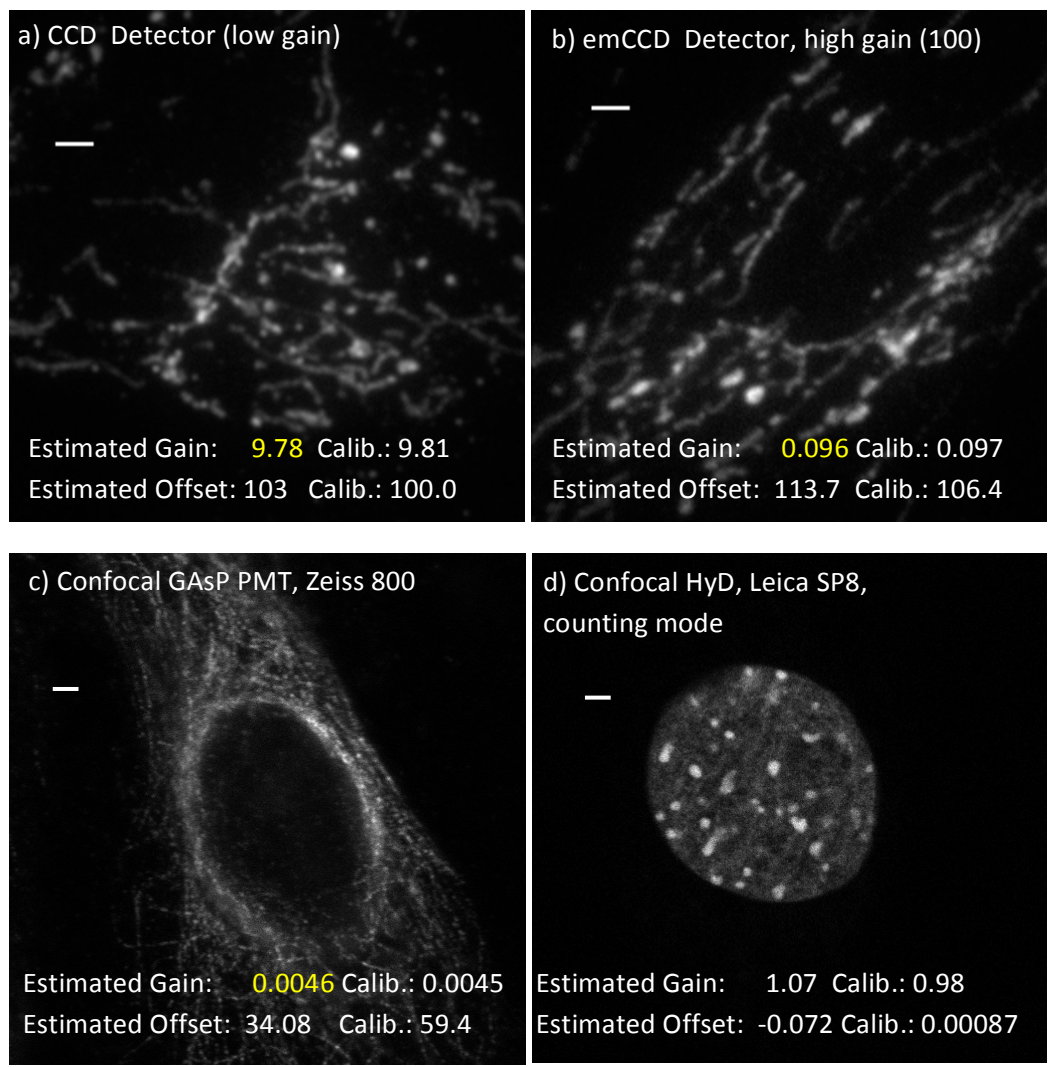

**Figure 2: Gain and offset estimation for various modes of acquisition**. Estimated gain and offset as determined for each image and the comparison to standard calibrated values. Panels a) and b) show mitochondria MuntJac AF555 cells. Panel c) displays labelled actin filaments (Alexa 568) and panel d) a DAPI-labelled cell nucleus of embryonic mouse fibroblasts.  (all scale bars 2 µm)

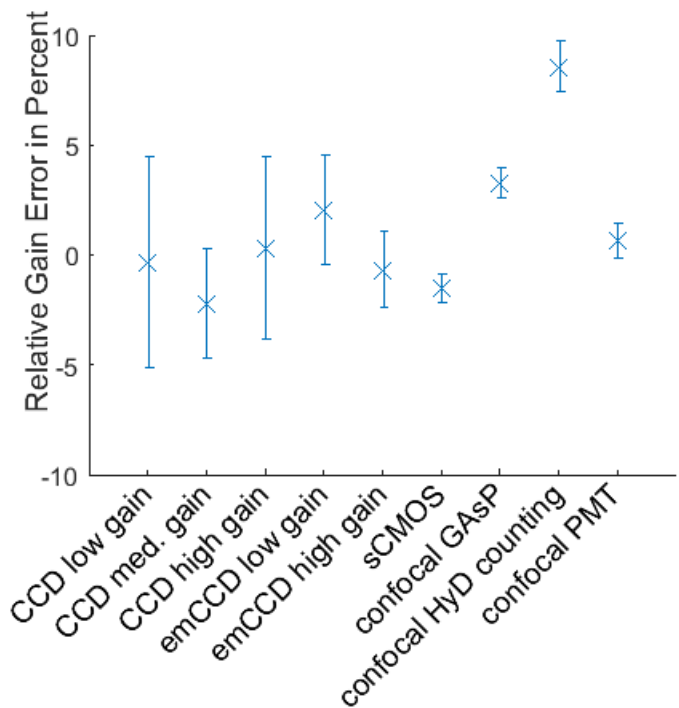

**Figure 3: Performance of single image gain determination on various detectors.** Relative error of the gain estimation from single image gain determinations compared with a standard "ground truth" calibration obtained from 20 smooth and 20 dark images each. The gain values varied from 0.1 (low gain CCD), 1.0 (confocal HyD counting) to almost 10 (high gain emCCD). Error bars indicate one standard deviation. See **Supplementary Figure 5** for example images for all imaging modes.

# Online Methods: Calibrating photon counts from a single image

Rainer Heintzmann[1,2*], Peter K. Relich[3], Robert P.J. Nieuwenhuizen[4], Keith A. Lidke[3], Bernd Rieger[4*]

[1]Leibniz Institute of Photonic Technology, Jena, Germany
[2]Institute of Physical Chemistry and Abbe Center of Photonics, Friedrich Schiller University, Jena, Germany
[3]Department of Physics and Astronomy, University of New Mexico, Albuquerque, NM, USA
[4]Quantitative Imaging Group, Department of Imaging Physics, Delft University of Technology, Delft, The Netherlands

## Experimental setups and methods:

**Figure 1**: Calibration images were produced by illuminating an opaque edge mounted on a microscope slide (VWR #48312-003) using a halogen lamp (Olympus U-LH100L-3). The sample was defocused to produce a smooth gradient and 100 images of the same pattern were acquired. Exposure times for each complete data set ranged between 15.7 ms to 51.2 ms. For each data set, a series of 100 background images were collected with the microscope camera shutter closed at the same exposure time.

**Figure 2**: For standard gain calibration 20 images were used in all panels. A commercially available test slide consisting of fluorescently labeled MuntJac skin fibroblast cells (Invitrogen F36925) was imaged using an EMCCD camera (Andor iXon 897) with conventional (panel A) or with various EM (panel B) gain settings: 20, 50, 100, and 200 and pre-amplifier gain settings: 1×, 2.6×, 5.2×. The microscope setup was an Olympus IX71 with an Olympus 150× 1.45 NA objective, 561 nm laser (Coherent Sapphire LP) excitation, 575 nm dichroic and a 600/30 nm band pass emission filter. The resulting back projected pixel size was 106.7 nm. The exposure time for all EMCCD data sets (panel B) was 20 ms. When the camera was operated with the EM gain disabled to act as a conventional CCD detector (panel A) the exposure time was 50 ms. Panel C shows an Alexa-phalloidin stained cell, imaged using a Zeiss LSM 800 confocal microscope using a Plan Apochromat 63× 1.4 Oil objective and GAsP detector. The pixel pitch was 82.5 nm. Imaged in the AF 568 channel. Panel D displays a DAPI stained cell nucleus imaged using a 100x 1.4 Oil HC PlanApo CS2 objective and the photon counting mode of the HyD detector of a Leica SP8 microscope. The pixel pitch was 79.6 nm.

**Figure 3**: This comparison of various detectors and imaging modes used the following settings: All CCD data was taken using various settings of an Andor iXon 897 (Andor, Belfast, U.K.) emCCD camera. For "CCD low gain", "CCD med gain" and "CCD high gain", the non-EM readout register was used and the pre-amplifier gain was set to 1, 2 and 3 respectively. The calibrated gain results were 9.8, 3.9 and 1.8 respectively. "emCCD low gain" and "emCCD high gain" used the electron multiplying register with EM gains set to 20 and 100 at pre-amplifier gain of 2 and 3 respectively. Calibrated gain results were 0.88 and 0.097 respectively. "sCMOS" corresponds to a Hamamatsu Flash 4.0 V2 (calibrated gain: 0.44). "confocal GAsP" used a Zeiss confocal LSM 800 Airy Scan with a GAsP detector in 16 bit mode (calibrated gain: 0.0045). "confocal HyD counting" refers to a Leica SP 8 confocal microscope using the HyD detector in photon counting mode (calibrated gain: 0.98). Maximal counts were about 90 ADU. "confocal PMT" used a Zeiss 800 microscope with a standard PMT detector (calibrated gain: 1.1). A value of $k_t$ =0.8 was used in the single-image based gain estimation algorithm and the *k*-axes cross was avoided as described.

## Derivation of the gain estimation from out-of-band information of a single image only:

We model the image formation as

(2) $$I(x) = g[B(x) + n_p(x)] + n_r(x) + O_{zero}$$

With *x* denoting the spatial detector coordinates, *B* the expected photon count at the detector, the photon noise $n_p$, the readout noise $n_r$ and the offset of the AD converter $O_{zero}$. Because noise is assumed to have a mean of zero and the noise sources can be assumed to be statistically independent, the expected value of *I(x)* is simply $g\,B + O_{zero}$ and the variance of *I* over time is equal to

(3) $$var(I) = g^2 var(n_p) + var(n_r) = g^2 B + \sigma_r^2$$

where $\sigma_r$ is the standard deviation of the readout noise. Our key idea is to extrapolate the expected variance of *I* from the average power at spatial frequencies *k* beyond a threshold frequency $k_t$ above the cut-off frequency $k_{cut}$ (out-of-band energy) of a single image because in this region all spectral power is entirely due to noise. Therefore the average spectral power of the noise *var(I)* at those frequencies is equal to the average spectral power of the noise at all spatial frequencies:

(4) $$\frac{1}{M}\sum_x var(I) = \frac{1}{M}\sum_x g^2 n_p^2 + n_r^2 = \frac{1}{M}\sum_k g^2 |\widehat{n_p}|^2 + |\widehat{n_r}|^2 = \frac{1}{T}\sum_{|k|>k_t}|\hat{I}|^2$$

where *M* is the total number of pixels and *T* is the number of pixels in the Fourier transform of *I* being above the threshold frequency $k_t$, the hat denotes the Fourier transformation and the summations over *x* and *k* denote the sum over pixels in respectively the spatial and Fourier domain. In the second step of equation (4) we used Parseval's relation (conservation of signal energy) to convert the sum over all pixels in the spatial domain into a sum in the Fourier domain. By combining all the above, the relationship between the variance and average measured intensity can be expressed as

(5) $$\frac{1}{T}\sum_{|k|>k_t}|\hat{I}|^2 = g\frac{1}{M}\sum_x I + \underbrace{(\sigma_r^2 - gO_{zero})}_{V_0}$$

or alternatively

(6) $$\frac{1}{T}\sum_{|k|>k_t}|\hat{I}|^2 = g\left(\frac{1}{M}\sum_x I - \underbrace{\left(O_{zero} - \frac{\sigma_r^2}{g}\right)}_{O_{no-noise}}\right)$$

Equations (5-7) are equal to equation (1) as given in the main text. Clearly the gain *g* can be found using equation (6) if the image offset $O_{zero}$ and the readout noise $\sigma_r$ are known:

(7) $$g = \frac{M\sum_{|k|>k_t}|\hat{I}|^2}{T\sum_x(I - O_{no-noise})}$$

with the definition of $O_{no-noise}$ as stated in equation (6), being the theoretical intensity value at which zero variance would be expected.

However, the offset is most often not known whereas the detector manufacturer specifies the readout noise (in most cases). Our algorithm first estimates $O_{no\text{-}noise}$ exploiting the linearity of the mean to variance relationship as detailed below to then compute the gain also in cases of unknown image offset.

## Gain estimation with an unknown image offset

We subdivide the image into tiles (e.g. 3x3 sub-image, the tiles) and plot for all tiles the out-of-band estimated mean noise energy $\frac{1}{T}\sum_{|k|>k_t}|\hat{I}|^2$ versus the measured mean intensity $\frac{1}{M}\sum_x I$. This gives 9 data points on the variance versus mean intensity plot and we estimate $O_{no-noise}$ by a linear regression on the data.

From this we then obtain the gain by computing the mean noise variance and mean intensity from the overall image as stated in equation (6). With this algorithmically determined $O_{no-noise}$, either the detector offset $O_{zero}$ or the readout noise $\sigma_r^2$ can be obtained if the other respective value is known a priory.

See Supplementary Information for a further discussion on the implementation details and fitting.

# Supplementary Information: Calibrating photon counts from a single image

Rainer Heintzmann[1,2], Peter K. Relich[3], Robert P.J. Nieuwenhuizen[4], Keith A. Lidke[3], Bernd Rieger[4]

[1]Leibniz Institute of Photonic Technology, Jena, Germany
[2]Institute of Physical Chemistry and Abbe Center of Photonics, Friedrich Schiller University, Jena, Germany
[3]Department of Physics and Astronomy, University of New Mexico, Albuquerque, NM, USA
[4]Quantitative Imaging Group, Department of Imaging Physics, Delft University of Technology, Delft, The Netherlands

## Simulation Results

To obtain information on the quality of the single-image-based gain and offset calibration procedure a series of simulation has been performed. We used two different types of images both 512×512 pixels 1) a resolution target and 2) a picture of Albert Einstein (compare Figure S1d) and S2d)), at an offset of 100 ADUs, where the ground truth of the gain is known.

These images were convolved with a PSF with its cutoff limit $k_{cut}$ at 2/3 of the Nyquist limit. A variety of maximally expected photons ($10,…,10^5$) and a number of readout noise contributions (1, 2 and 10 photons) were simulated. For each of these conditions, 100 simulations have been performed. For each individual simulated image the gain and offset where determined by our algorithm using $k_t = k_{cut}$ in the algorithm, and supplying the nominal readout noise to it. All simulated and experimental data is based on the 3×3 tile offset estimation method as presented below. The results were then compared to the ground truth (nominal gain and photon count obtained via detector calibration experiments) and summarized in Figure S1, S2 and S3. Here the midpoint of a "relative gain error in percent" entry is computed by $\frac{\sum_{i=1}^{100} \hat{g}_i - g_{nominal}}{g_{nominal}} \times 100\%$, where $\hat{g}_i$ is the estimated gain by our algorithm in simulation number $i$. The error bars refer to plus and minus the corresponding relative value of the standard deviation.

There are 4 different relevant spatial frequencies for the simulations:
- $k_{cut}$    The optical cut-off frequency of the OTF.
- $k_t$    The threshold frequency for computation of spectral power density of the noise $V_{HF}$.
- $k_{Nyquist}$   The required Nyquist sampling frequency given $k_{cut}$, i.e. $k_{Nyquist} = 2k_{cut}$.
- $k_{samp}$    The actual sampling frequency.

Ideally we have $k_{samp} = k_{Nyquist} >= 2 k_{cut} = 2 k_t$. But in reality these relations might differ slightly. In practice the actual sampling might often differ from the strictly necessary one. Our algorithm can only vary $k_t$ as the rest is determined upon acquisition. However, as different relative values of these frequencies influence foremost the correct estimation of the spectral power density of the noise $V_{HF}$, we investigate the performance of the algorithm with respect to several cases. We created a slightly oversampled test image with $k_{cut}$ =2/3* ½ $k_{Nyquist}$ and we specify $k_{cut}$ as the fraction to the matching Nyquist sampling, i.e. here as $k_{cut}$=2/3. We specify $k_t$ also relative to ½ $k_{Nquist}$, i.e. $k_t$=2/3=$k_{cut}$ will only consider pure noise for the estimation of $V_{HF}$. Lower values of $k_t$ will consider signal into computing $V_{HF}$ which is in effect similar to spatially undersampling the image.

Observations from simulations on the gain estimation:
- The higher the readout noise the worse the gain estimate.
- Higher photon counts give better results in case of high readout noise.
- The higher the gain, the higher the variance in the estimation of the gain. However, the relative gain error was essentially independent of the gain. A small positive constant bias of about 1% was observed, which stayed below (resolution target) or was comparable to (Einstein) the standard deviation of the estimation.
- Readout noise of 10 at 10 maximally expected photons gave rise to very large variances. However, the mean estimate is still within 10-20% error (compare SFig 1c) and 2c)).
- $k_t$ = $k_{cut}$ yields the best results. For $k_{cut}$=2/3 there is plenty of out-of-band noise present in the pixelated image due to the oversampling. Even for small undersampling 10-20% e.g. 0.9$k_t$=$k_{cut}$ we still have acceptable results.
- Gain estimation errors with choosing $k_{cut}$ below the cut-off limit, i.e. $k_{cut}$=2/3 and $k_t$ = 0.5, remains acceptable with errors below 10% except for $10^5$ expected photons or more, as more and more signal is wrongly assumed to be noise in the computation for $V_{HF}$. If we decrease $k_t$ further below $k_{cut}$ lower photon counts (<$10^3$) already show an increased error as expected (data not shown).

Instead of investigating directly the relative offset estimation error as this value cannot be reliably obtained independently from the measured signal, we instead evaluate the performance of the algorithm to estimate the effective photon count. For data processing purposes the physical offset is often irrelevant as the offset-corrected data converted to effective photons is used for any further analysis such as maximum likelihood deconvolution. The term "effective photons" refers to a signal with the noise variance being equal to the mean value. Therefore we investigate the performance for effective photon count $N$

$$N = \frac{S - O_{no-noise}}{g}$$

where $S$ is the image value in ADU, $O_{no-noise}$ is the ADU offset (ADU value at zero effective photons) and $g$ the gain. By propagation of error we find the variance in the photon count estimation

$$(\Delta N)^2 = \left(\frac{\partial N}{\partial g}\right)^2 (\Delta g)^2 + \left(\frac{\partial N}{\partial O}\right)^2 (\Delta O_{no-noise})^2 =$$
$$= \left(\frac{N}{g}\right)^2 (\Delta g)^2 + \frac{1}{g^2} (\Delta O_{non-noise})^2$$

Here Δ*g* and Δ*O*$_{no\text{-}noise}$ are the standard deviation of the estimate of the gain and offset respectively, which are determined from the simulations over 100 runs. As the simulated photon count spans 5 orders of magnitude we plot the relative error Δ*N/N*·100% in Figure S3. The variation in the gain estimate dominates the error in the photon count in most cases.

As seen from these results (errorbars in Figure S1 and S2), the mean relative error in the gain estimation is independent of the actual gain, confirming that the algorithm is invariant to multiplicative factors. Only in the presence of substantial readout noise an influence of a multiplicative factor is seen. It is further observed that the bias stays below 1% for all cases but when simulating with a maximum number of 10 expected photons.

The only case, which leads to an unacceptable high variance in the estimation, is for 10 photons readout noise at a maximal signal level of 10 photons. An example image for this case is shown in Figure S4. Interestingly even in this case we can obtain an acceptable result (standard deviations ~10-20% for the lower gain values), if the threshold frequency $k_t$ supplied to the algorithm is reduced from 2/3 to 0.2. The reason is that at such noise levels the signal is completely lost in the noise already at frequencies far below the actual cut-off limit. A reduced threshold limit $k_t$ then helps to estimate the noise more robustly. However, at good SNR (high photon counts), such a lowered $k_t$ leads to unwanted biases as signal energy is falsely accounted for as noise energy.

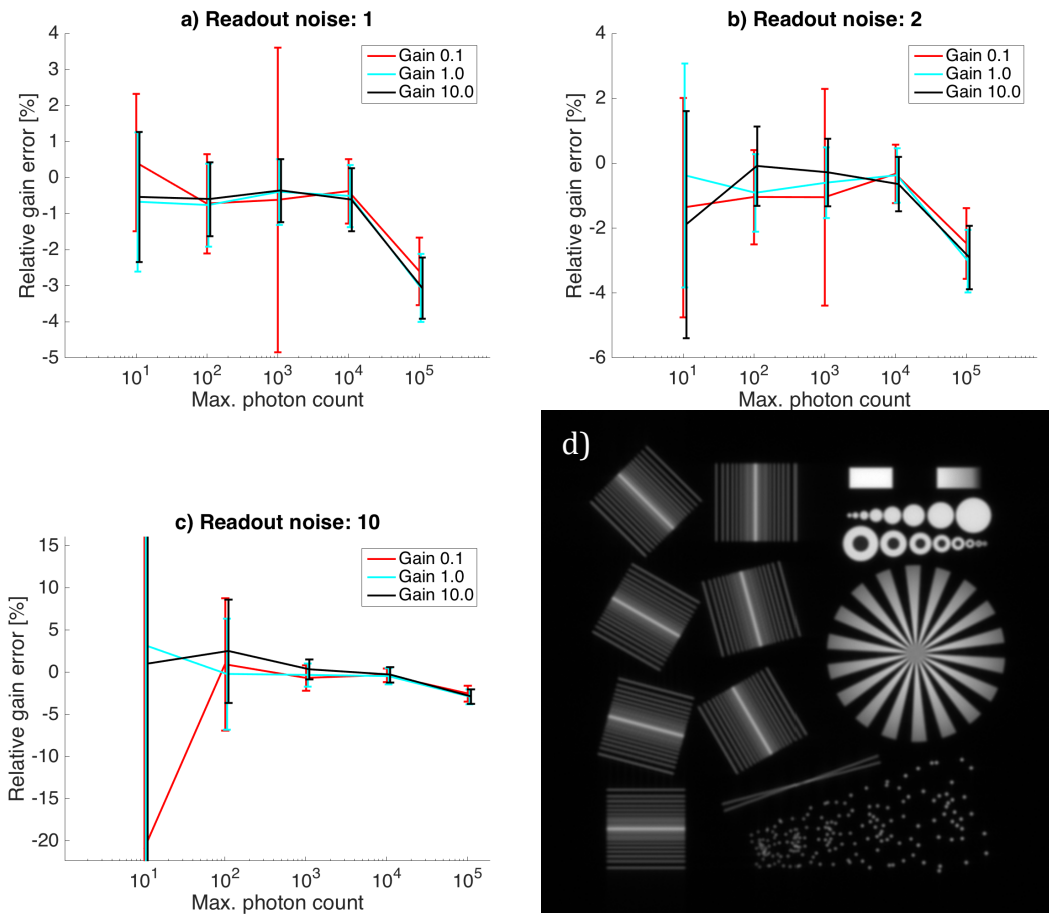

**Figure S1:** Performance of the single image gain estimation on a simulated resolution target as shown in d). Panels a, b, c) show the relative gain estimation error for respectively different levels of readout noise with standard deviation of 1, 2 and 10 photons as a function of maximal photon count (10,...$10^5$) for three gain levels. Markers indicate mean error over 100 runs and error bars one standard deviation of the 100 runs. In panel c) the standard deviation of the results for 10 expected photons in the maximum which ranged up to 600% is not shown. For 100 photons only for gain 10 it is outside the display range. Panel d) shows the simulated resolution target (512×512 pixels). Parameters $k_t = k_{cut} = 2/3$.

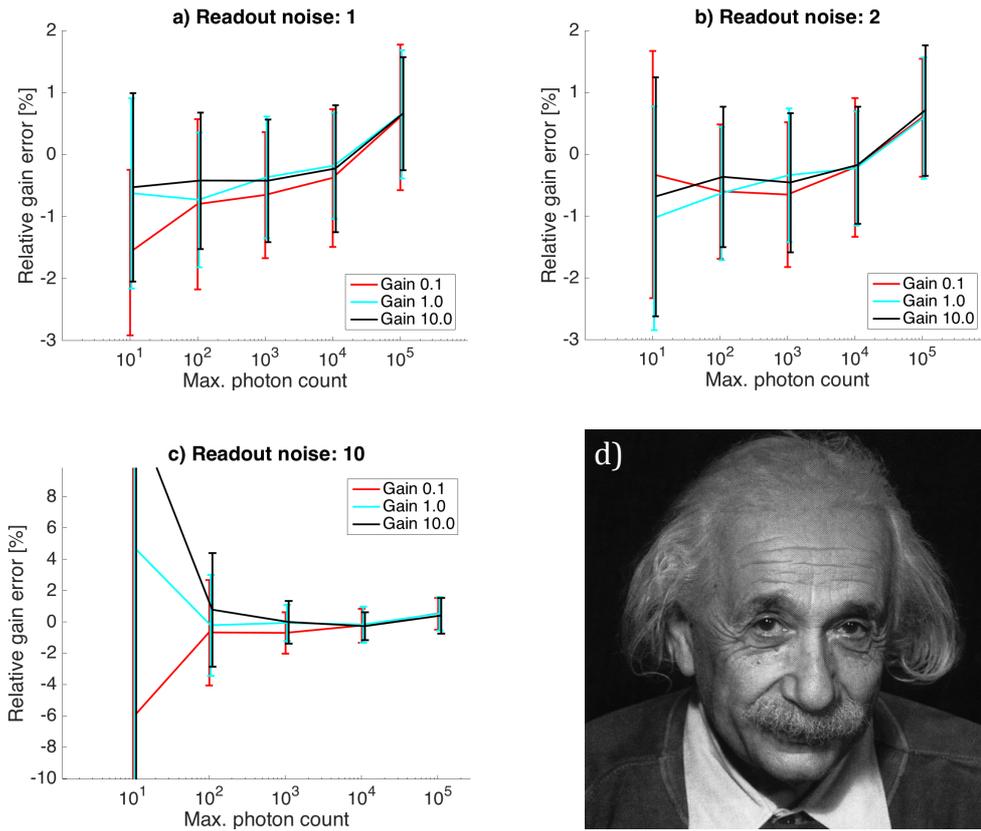

**Figure S2** Performance of the single image gain estimation on Einstein (512×512 pixels) as shown in d). Panels a, b, c) show the relative gain estimation error for different levels of readout noise with standard deviation of respectively 1, 2 and 10 photons as a function of maximal photon count (10,…$10^5$) for three gain levels. Markers indicate mean error over 100 runs for different simulated gain values. Parameters $k_t$ = $k_{cut}$ = 2/3.

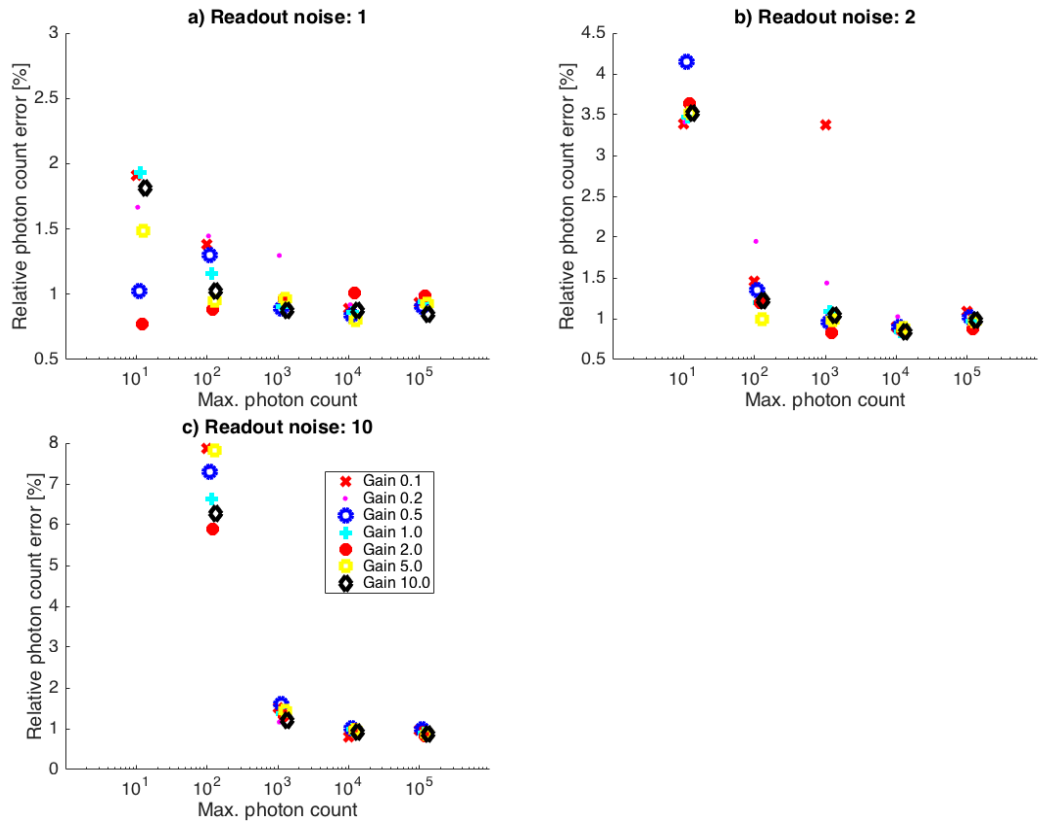

**Figure S3:** Estimated relative error in the photon count from the single-image estimation using the Einstein object as simulation input. Panels a, b ,c) show the relative photon count estimation error for different levels of readout noise with respective standard deviations of 1, 2 and 10 photons as a function of maximal photon count $(10,…10^5)$ for 7 gain levels. In panel c) we omitted the results for 10 expected photons where estimation breaks down and errors above 100% were found.

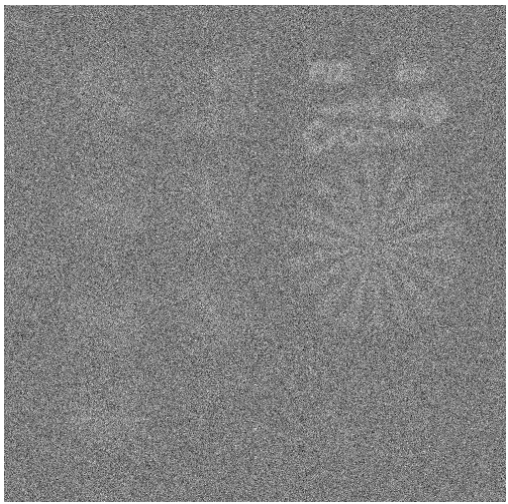

**Figure S4:** Worst SNR in simulations for readout noise 10 photons and 10 photons expected maximum photon count.

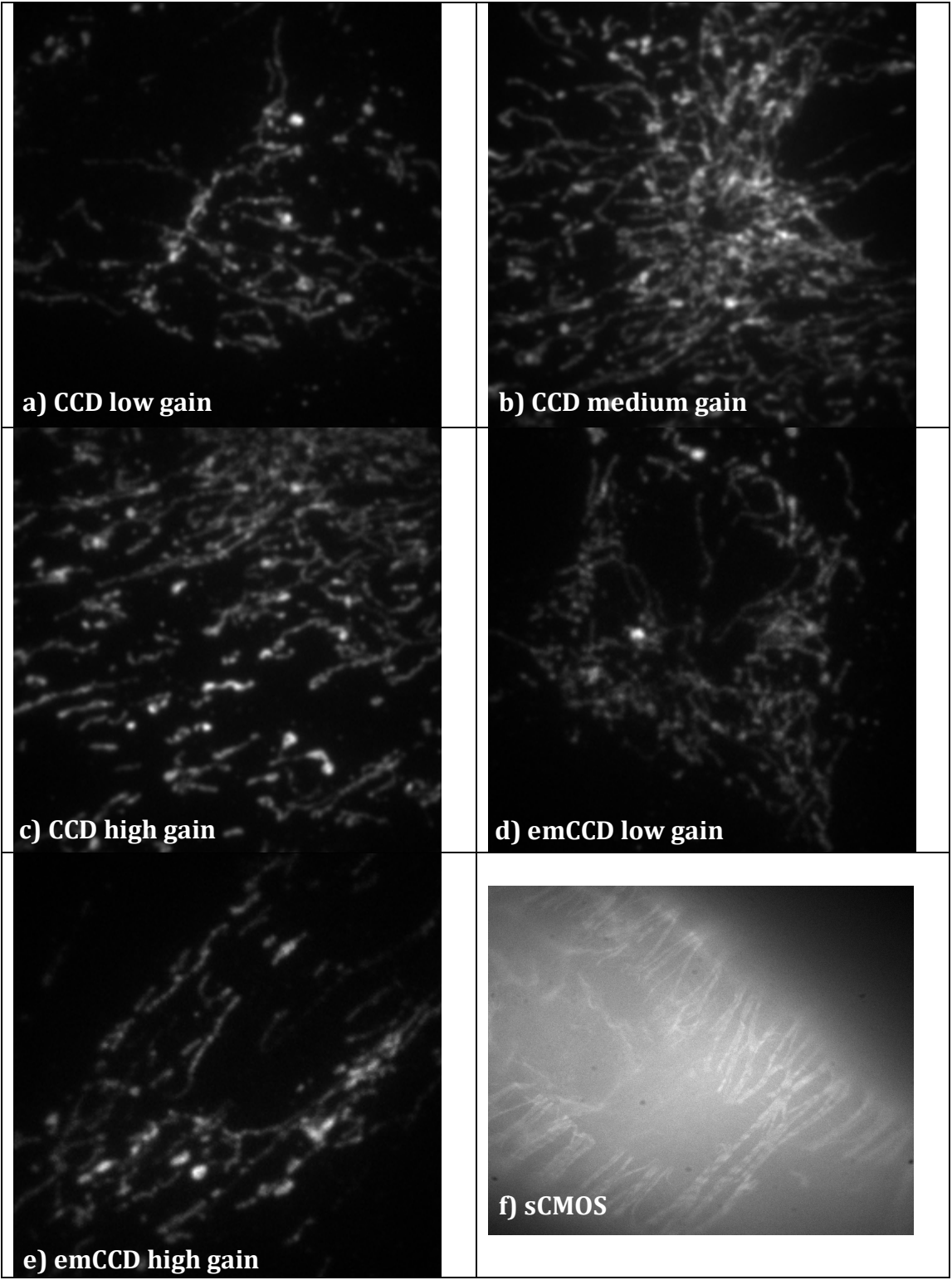

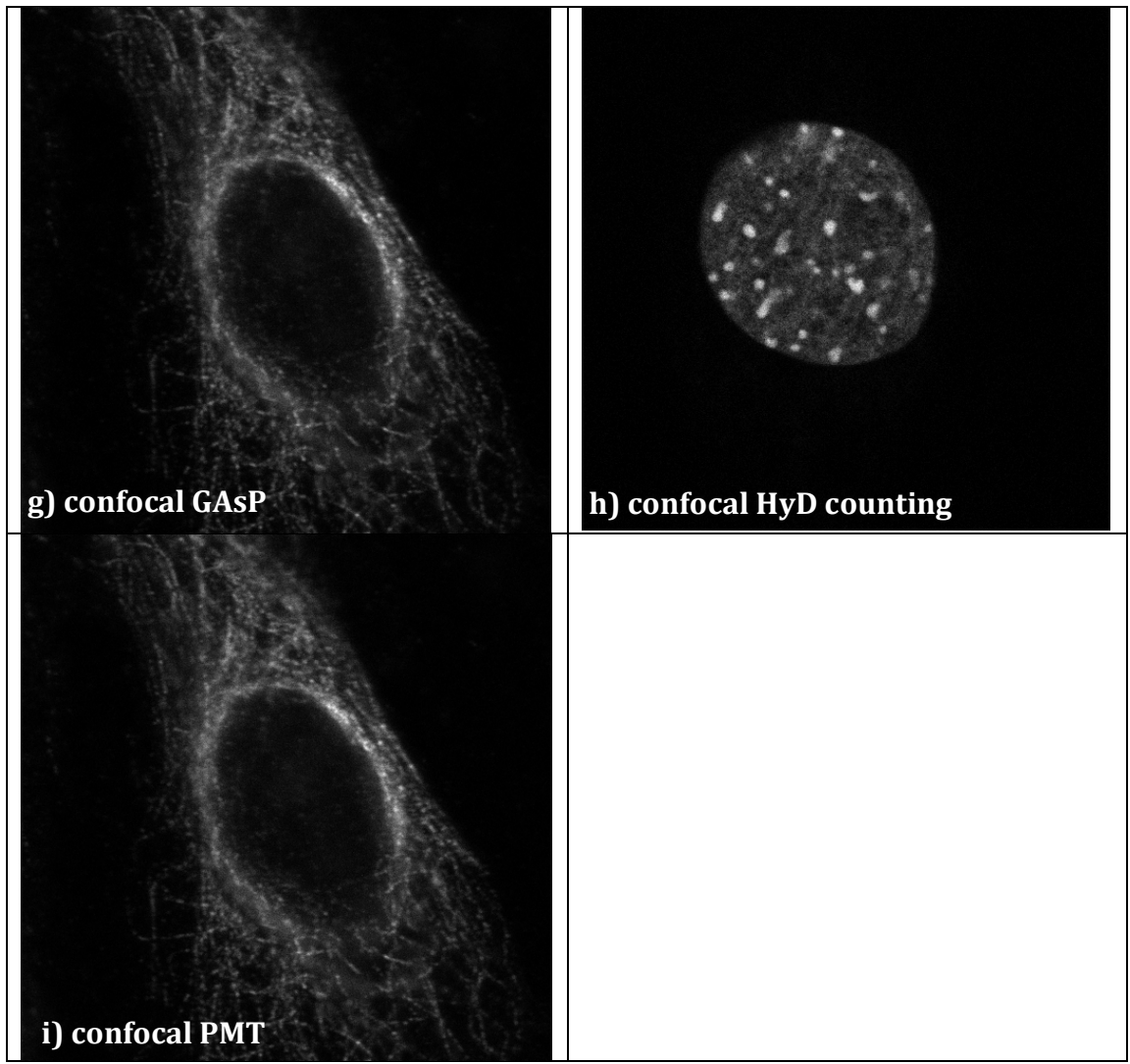

**Figure S5: a)-i)** Example images used for the single image photon quantification on experimental data of various instruments and detector types. Respective results of the estimation are shown in Figure 3.

## Traditional Gain estimation from a series of exposures

Here we describe the standard gain calibration procedure according to [vanVliet1998, Lidke2005].

Two series of images, each typically about 20 images, are acquired a dark series with no light exposure and a series with a smooth intensity distribution (gradient sequence as exemplified in the main text, Figure 1, left inset). The intensity distribution should capture the full linear dynamic range of the detector without saturation. For the upper limit we advise about half the maximum of the linear working range of the device, the full well capacity of the camera or the maximum of the ADU converter range. Such a distribution can be obtained by finding a bright edge or isolated structure and defocussing it. It is important to ensure that both series are acquired under identical settings, temperatures and exposure times.

These series are then processed by first determining the offset ($O_{zero}$) as the mean value of all dark frames (see Fig. 1). Such an estimate is usually precise since images contain many pixels and this is averaged over many frames. Of course such a background series can also prove useful to estimate offset-drift effects by plotting the mean of each frame as a time series, or residual noise sources by averaging the magnitude of all individual 2D Fourier transforms, where even weak periodic fluctuations often show up as clearly visible peaks. Then from the sequence of smooth images, the mean and the variances are calculated. To this end the image is binned into regions of intensity bins determined from the average image. We use a standard of 100 intensity bins to obtain a robust estimate and account for drift effects that could influence pure pixel based estimation. The gain $g$ is obtained from the slope of the fit of $var(S) = g\,(S - O_{no-noise})$ to the plot of the mean variance over mean intensity (see Fig. 1).

The signal output $S$ in ADU comes from the detector in response to an input of $N$ effective photons which are obtained as $N = (S - O_{no-noise})/g$. Note that for Poisson statistics it holds that $var(N) = N$. The fit is a reweighted least square method with the square of the variance and the number of averaged variance values in each bin predicting the variance of the variance. From the slope of such a plot, the gain can be estimated as given in Figure 1. With this knowledge the mean of the variance projection of the dark series can be converted to the readout noise of the device in electrons.

## Description of the single image gain estimation algorithm

The mean spectral energy of the noise ("out-of-band energy") is computed as depicted in Figure S6 by $\frac{1}{T}\sum_{|k|>k_t}|\hat{I}|^2$, where $T$, is the number of Fourier-space pixels beyond the algorithmic cut-off frequency $k_t$ compared to the area of the full image. The threshold frequency $k_t$ is typically equal or larger than the optical cut-off frequency $k_{cut}$. For a discussion on practical values of $k_t$ see discussion on the Simulation Results. The Fourier transformation is computed using the discrete Fourier transformation (DFT) as outlined in the implementation details below.

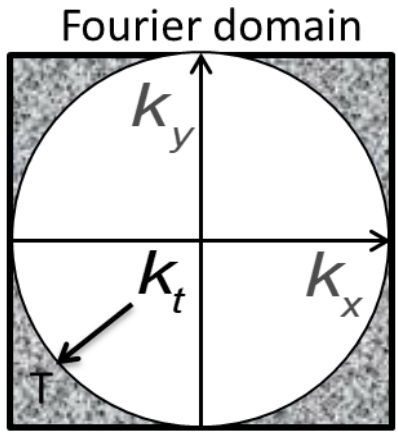

**Figure S6:** Spectral energy in the out-of-band region in the Fourier domain of the image. T is the number of pixels beyond the algorithmic threshold frequency $k_t$.

## Gain estimation and image offset

To be able to relate measured values to noise prediction, we need to determine the variance offset ($V_o$) at zero readout value. As detailed in the Online Methods we tile the input image into a number of sub-images or tiles. It is not of the essence how many or how large the tiles are for the estimate. We chose 3×3 tiles, which worked well in simulations and was also used for all computations of experimental data. The workflow is depicted in Figure S7. Increasing the number of tiles gives more data points on the variance versus sum intensity plot, but each data point has a higher uncertainty as it is computed from a lower number of pixels. In addition, the different tiles need to cover a range of mean intensities to give a good scatter on the plot and reducing the tile size too much, does not foster this requirement. Once the variance offset $V_o$ is found by reweighted linear regression (as described below) the gain is estimated as follows.

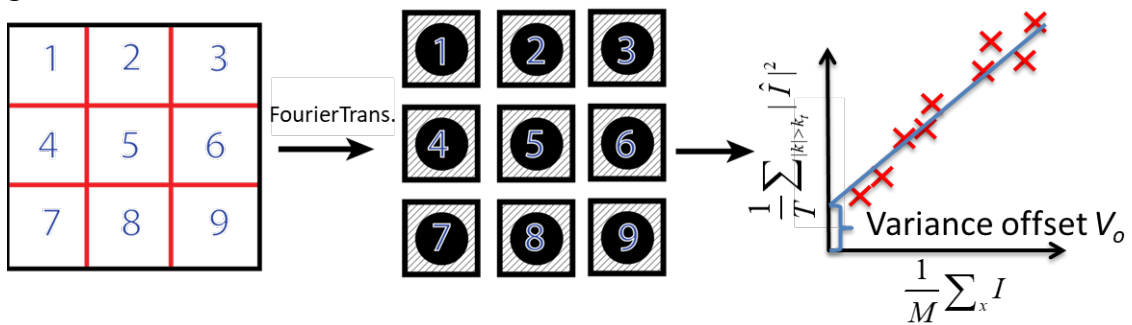

**Figure S7:** Workflow to enable gain estimation from a single image in the case of unknown image offset. The input image is divided into 3×3 sub-images. For each sub-image the Fourier transformation is computed and the mean out-of-band energy of the noise is measured. These values are plotted against the mean intensities in the tiles (including the unknown offset). From a reweighted linear regression we find the variance offset $V_o$, which is finally used to compute the gain.

We plot this sum against the variance due to noise in each tile corrected for the fraction of the area. The variance offset at zero readout value $V_o$ is then used to find the overall gain. To this end we compute again $\sum_{|k|>k_t}|\hat{I}|^2$ and $\sum_x I$ but now on the

whole input image and finally obtain the gain as
$$g = \frac{M \sum_{|k|>k_t}|\hat{I}|^2 - V_o}{T \sum_x I}.$$

To relate measured values not only to a noise prediction but to actual photon counts, we need to supply either the readout noise (in e⁻ RMS) $\sigma_r$ or the image offset ($O_{zero}$ in ADU) to our algorithm. We obtain the variance of the readout noise as $V_{read} = \sigma_r^2$. If the readout noise is supplied, the image offset $O_{zero}$ is obtained after the gain $g$ is found, and $V_o$ from the regression as $O_{zero} = \frac{V_{read}}{g} - \frac{V_o}{g}$.
To convert the image from ADU to effective photon counts the offset $O_{zero}$ is first subtracted before we divide by the gain. Note that for higher photon counts, the readout noise estimate does not need to be precise, still yielding good estimates of the gain and offset. For images with very low photon counts with unknown readout noise, it may be advantageous to rather estimate the offset from a dark area of that image.

### Weighted linear regression on variance vs. sum intensity
The obtained mean-variance data is fitted by a Poissonian reweighted least square approach to find the variance offset $V_o$ as giving in Figure S7. The errors in the $y$ values are given by the squares of the variances, divided by the number of pixels in the tile. The variances itself are just given by the mean $y$ values for Poisson statistics. The fit weights are then the inverse of this error and the slope and offset (as vector $\beta$) are found from $\beta = (X^t W^{-1} X)^{-1} X^t W^{-1} y$. Where $W$ is the covariance matrix of the weights (here the measurements are independent and the matrix is diagonal) and $X$ is the matrix of $x$ values given by sum of tile intensities.

### Implementation details
The discrete Fourier transformation unavoidably implies that the input signal is cyclic which gives rise to artifacts (false spatial frequencies) in the Fourier domain. To avoid this well-known problem typically a Tuckey, Hanning window or similar is applied to the image before Fourier transformation. However, applying any such window also disturbs the noise power spectrum and affects the single image gain estimation. Therefore we have chosen not to apply any window but to use mirroring of the input image and tiling it into a 2×2 array before Fourier transformation as displayed in Figure S8. This operation ensures matching boundaries and avoids zero-order discontinuities of the discrete Fourier transformation but leaves the noise distribution unperturbed.
Some detectors, such as sCMOS cameras, can exhibit fixed pattern noise and/or pixel dependent gain extending into the high-frequency region [Huang2013]. Diffusion and blinking during (confocal) scanned image acquisition or afterpulsing can also leads to similar effects, as for example exploited in RICS [Brown2008]. These can lead to potential problems of our method. Therefore we additionally chose to remove the spectral cross from the Fourier domain before estimation the spectral energy of the noise. That is, we null out a 3 pixel wide stroke on the Fourier axes next to nulling all spectral content $|k|<k_t$ as shown in Figure S6 and account for this in the fraction $f$.

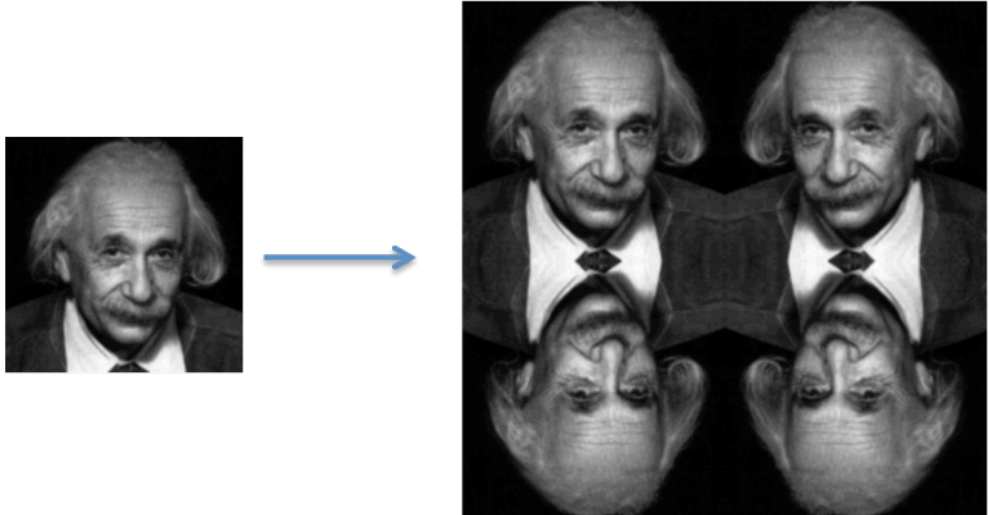

**Figure S8:** Symmetrizing an input image by mirroring to avoid boundary discontinuities when applying the discrete Fourier transformation.

Note: If the input image does not have an equal number of pixels along both dimensions and a square pixel size the discrete Fourier transform will result in non-isotropic pixels in Fourier domain and the physical cut-off frequency will not be a circle in the Fourier transformed image, but an ellipse.

Software can be downloaded from our website http://www.diplib.org/add-ons provide Matlab (The Mathworks, MA) code and an ImageJ (NIH, VA) plugin.

## Comparison with prior art on single-image based algorithms

Estimation of the combined Gaussian and/or Poissonian noise parameters from single images has been studied in the signal processing community [Immerkaer1996, Foi2008, Pyatykh2014]. The estimation strategies in the literature rely on finding the mean-variance intensity curves from either homogeneous image parts/blocks or after high-pass filtering and attempting to remove signal contribution from the image. For an extensive review see Colom et al. [Colom2014] and Pyatykh et al. [Pyatykh2014]. In other words assumptions are made on the typical properties of objects in the image. We approached the problem from the acquisition side. In stark contrast to these methods we only assume that the image is acquired by a band-limiting process (such as given by any optical imaging system) and that it was at least roughly sampled correctly. That means our approach is independent of the image content and structure. For estimation of the Poisson component of the noise the normal approximation of the Poisson distribution has been applied in the prior art (or the Anscombe transform was used). In our application, light microscopy and in particular fluorescence imaging, the detection process is largely shot-noise limited due to the low signal intensities, but also the very low readout noise of current (em)CCD and sCMOS detectors.

Gain estimation on the experimental data from Figures S5 with the available implementations by Pyatykh et al. [Pyatykh2014] did typically result in errors larger than 100% (and a much larger spread) when compared to the gold standard calibration and their implementation got stuck in endless loops for some of the experimental and simulated datasets (Figure S9). The algorithm has 15 free parameters where we used the ones set in the software provided by the authors. Potentially fine-tuning the parameters could solve the problem of the algorithm getting stuck but we did not pursue this effort as our fine-tuning 15 parameters seemed fruitless compared to our very easy routine. For the CCD med. gain and emCCD high gain mode the code did not return a result and we skip these data points.

The performance of the implementation of Azzari2014 et al. [Azzari2014] was a bit better with relative errors of 10-600% and smaller spread. However, compared to our results with a relative gain error of on average 1.4% (0.8% median) (Figure 3) the Azzari2014 method is still about two orders of magnitude worse than our approach on experimental data with a mean relative error of 181% (66% median), as seen in Figure S9. For the best estimations of the Azzari method, namely the confocal GAsP case and the confocal HyD counting case, the relative error values are 9.6+-1.6% and 9.6+-5.1% respectively, which are also still worse than our estimations of 5.3+-0.7% and 8.4+-1.2% (compare Figure 3).

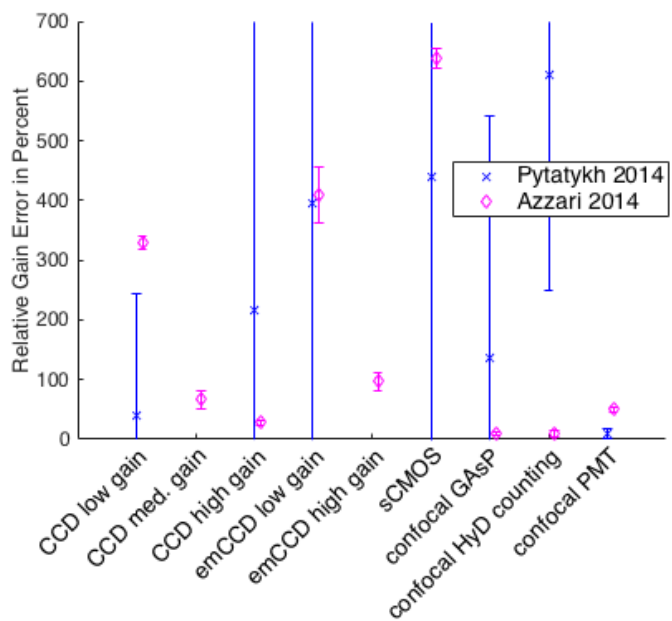

**Figure S9: Performance of algorithm by Pytatykh et al. and Azzari et al. for gain determination on various detectors.** Relative error of the gain estimation from single image gain determinations compared with a standard "ground truth" calibration obtained from 20 smooth and 20 dark images each. The gain values varied from 0.1 (low gain CCD), 1.0 (confocal HyD counting) to almost 10 (high gain emCCD). Error bars indicate one standard deviation. See Figure S5 for example images for all imaging modes.

Estimation on simulated data was for both investigated algorithms much worse than our algorithm. On the image of the resolution target we show in Figure S10 the relative gain estimation errors over 50 noise realizations. It is evident that especially the algorithm by Azzari et al. results in very poor estimates with 50-100% relative errors whereas the algorithm of Pyatyth et al. is in the range of 0-50%. Please note that the results are gain dependent. The standard deviation is omitted, as it is larger than the relative errors. Please also note that the photon count is restricted to $10^4$ photons as for larger counts the code got stuck like for the experimental data in an endless loop.

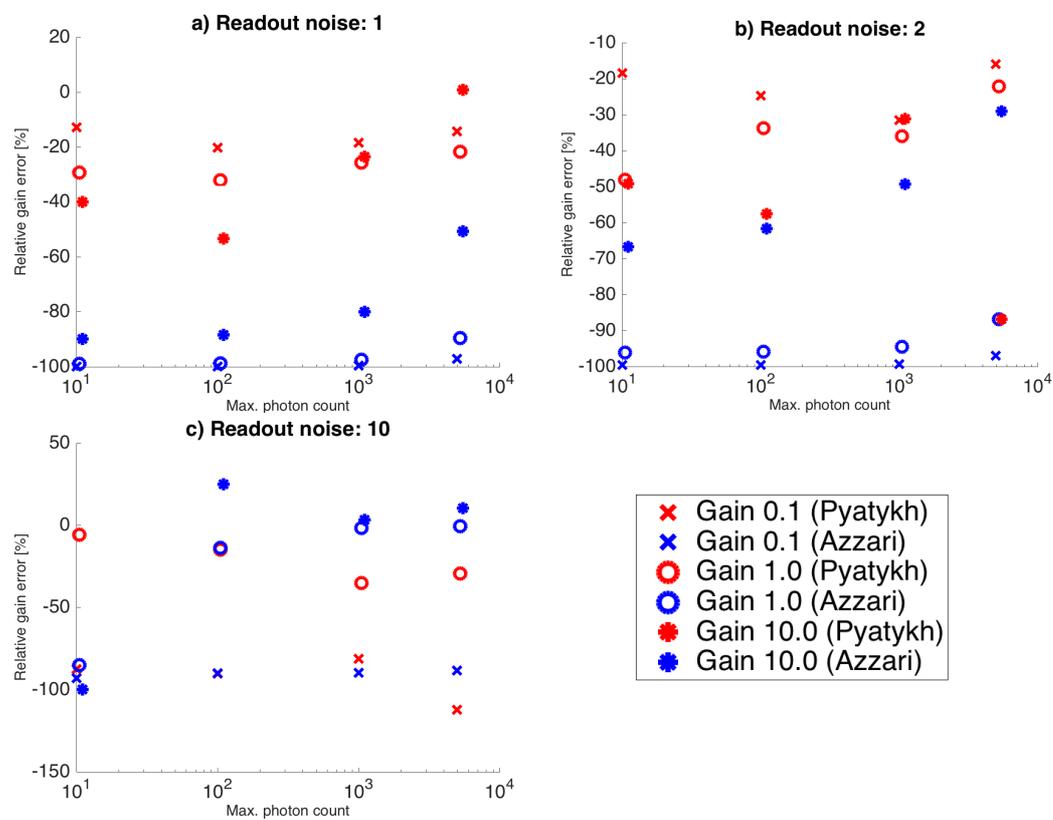

**Figure S10: Performance of algorithm by Pytatykh et al. and Azzari et al. for gain determination on simulations for the "resolution target" image.** Mean relative error of the gain estimation over 50 noise realizations for different amounts of added readout noise. Simulation settings are identical to Figures S1 and S2.

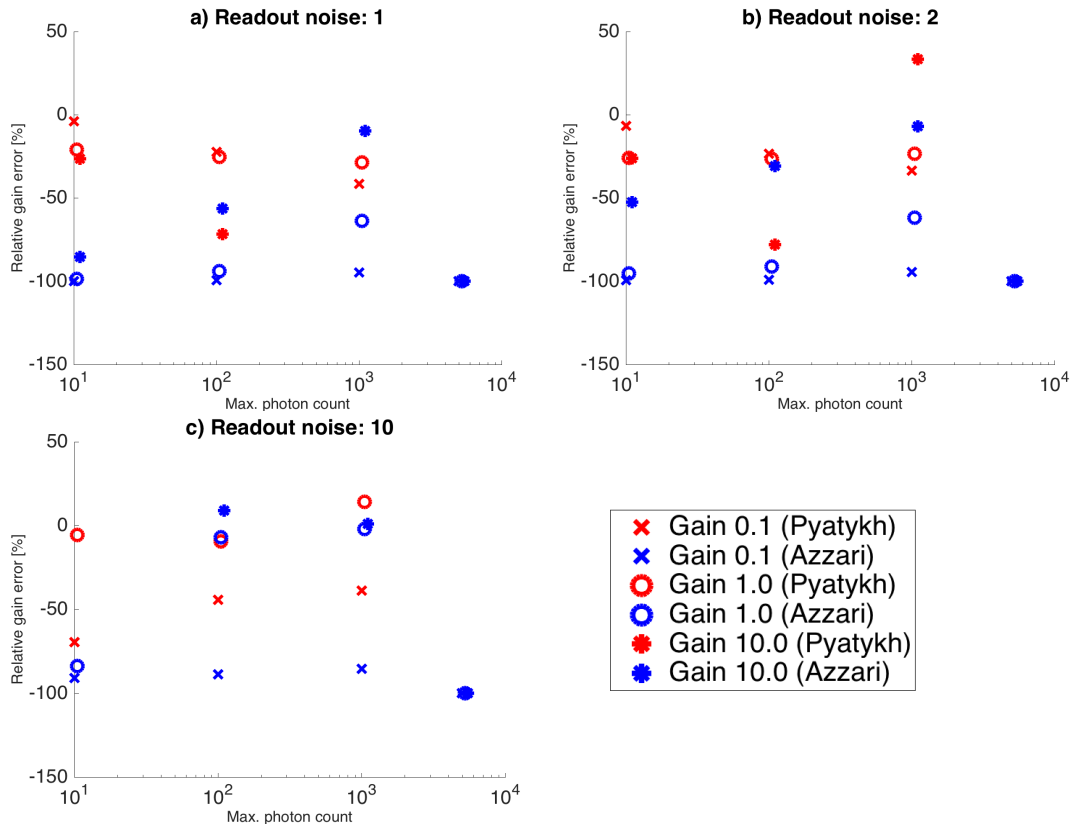

**Figure S11: Performance of algorithm by Pytatykh et al. and Azzari et al. for gain determination on simulations for the "Einstein" image.** Mean relative error of the gain estimation over 50 noise realizations for different amounts of added readout noise. Simulation settings are identical to Figures S1 and S2.

Our current algorithm assumes that the image does not contain saturated pixels, i.e. no values that are clipped to the maximum bit depth of the A/D converter. In (fluorescent) microscopy this issue is irrelevant as light levels are very low and students are taught in entry-level courses already as best practise to never saturate images by increasing any variable gain. Prior art suggest to be able to cope with such undesirable acquisition problem but failed in our hands already for properly acquired data.

We are also largely independent of the image content/structure as long as there are different image intensities present in different parts of the image otherwise the tiling approach might fail. That is also in contrast to earlier methods that have difficulties to deal with strong textured content.

Prior art also suggest being able to cope with undersampling issues. Our experimental and simulated data are Nyquist sampled or better, yet the aforementioned published algorithms show poor performance. Commercial software for integrated setups commonly points users to the required Nyquist sampling rate of $\lambda/(4NA)$. This allows the user to match it e.g. by adjusting the pixel size in a confocal system. For widefield acquisitions an objective with a higher magnification or an additional optical relay system can fix potential undersampling for a given camera pixel size. As indicated above our methods also tolerates some degree of undersampling (10-20%) especially for images of low SNR as the microscope OTF falls off to zero strongly towards the optical cut-off frequency and only a small fraction of the signal energy is then wrongly assumed as noise energy. Furthermore have slightly undersampled images still an out-of-band region in the corners of the Fourier-transformation, which can be used for our algorithm (see Figure S6).